\documentclass[twocolumn,amsmath,amssymb]{revtex4}
\usepackage{graphicx}
\usepackage{dcolumn}
\usepackage{bm}
\begin{document}

\title{Understanding the Missing Fractional Quantum Hall States in ZnO}
\author{Wenchen Luo and Tapash Chakraborty\footnote{Tapash.Chakraborty@umanitoba.ca}}
\affiliation{Department of Physics and Astronomy,
University of Manitoba, Winnipeg, Canada R3T 2N2}

\date{\today}
\begin{abstract}
We have analyzed the crucial role the Coulomb interaction strength plays on the even and odd denominator
fractional quantum Hall effects in a two-dimensional electron gas (2DEG) in the ZnO heterointerface.
In this system, the Landau level gaps are much smaller than those in conventional GaAs systems. The Coulomb 
interaction is also very large compared to the Landau level gap even in very high magnetic fields. 
We therefore consider the influence of higher Landau levels by considering the screened Coulomb 
potential in the random phase approximation. Interestingly, our exact diagonalization studies of the 
collective modes with this screened potential successfully explain recent experiments of even and odd 
denominator fractional quantum Hall effects, in particular, the unexpected absence of the 5/2 state
and the presence of 9/2 state in ZnO.

\end{abstract}

\maketitle
Discovery of the odd-denominator fractional quantum Hall effects (FQHE) in GaAs heterojunctions in 1982 
\cite{fqhe} and its subsequent explanation by Laughlin \cite{laughlin,book}, has remained the `gold 
standard' for novel quantum states of correlated electrons in a strong magnetic field. These effects 
also have been observed in `Dirac materials' such as graphene \cite{abergeletal,graphene_book,FQHE_chapter}.
and are expected to be present in other graphene-like materials \cite{silicene,germanene,Dressel}
with novel attributes. The FQHE states in monolayer and bilayer graphene were investigated theoretically 
\cite{FQHE_chapter,Monolayer,Bilayer1,Bilayer2} and experimentally \cite{FQHEGraphene,FQHEBilayer}.
For example, in bilayer graphene the application of a bias voltage results 
in some Landau levels (LLs) a phase transition between incompressible FQHE and 
compressible phases \cite{Bilayer1,Bilayer2}. The FQHE in silicene and germanene 
indicated that because of the strong spin-orbit interaction present in these materials as
compared to graphene, the electron-electron interaction and the FQHE gap are significantly modified 
\cite{Buckled}. The puckered structure of phosphorene exhibits a lower symmetry than graphene. 
This results in anisotropic energy spectra and other physical characteristics of phosphorene, both 
in momentum and real space in the two-dimensional (2D) plane \cite{Kou,LiuReview}. The anisotropic
band structure of phosphorene causes splitting of the magnetoroton mode into two branches with two 
minima. For long wavelengths, we also found a second mode with upward dispersion that is clearly 
separated from the magnetoroton mode and is entirely due to the anisotropic bands \cite{phospho}. 

In 1987, a discovery of the quantum Hall state at the LL filling factor 
$\nu=\frac52$, the first even-denominator state observed in a single-layer system \cite{willett}
added to the mystery of the FQHE. It soon became clear that this state must be different from the FQHE in 
predominantly odd-denominator filling fractions \cite{fqhe}. Understanding this enigmatic state has 
remained a major challenge in all these years \cite{jim_tilted,TC_even}. At this half-filled first 
excited LL, a novel state described by a pair wave function involving a Pfaffian \cite{Read,Bilayer2},
where the low-energy excitations obey non-Abelian exchange statistics, has been the strongest candidate.

The field of FQHE has now witnessed a very exciting development with the the observation of the
effect in high-mobility MgZnO/ZnO heterointerfaces \cite{zno,tsukazaki}. The odd-denominator
fractional states such as $\nu=\frac43, \frac53$ and $\frac83$ were observed here with indications of the
$\nu=\frac25$ state in the extreme quantum limit. Soon after, the even-denominator states, such as
$\nu=\frac32$, and $\frac72$ were also observed \cite{falson}, but surprisingly, the most prominent
even-denominator state of the GaAs systems, the $\nu=\frac52$ is conspicuously absent in the ZnO system.
The system of 2DEG in ZnO is unique as compared to that in GaAs. In the case of GaAs-based 2DEG, the
LL gap is large compared to that for the Coulomb interaction ($e^2/\epsilon\ell$, 
where $\epsilon$ is the dielectric constant and $\ell=\sqrt{\hbar /eB}$ is the magnetic length with 
a magnectic field $B$). However, in a ZnO heterosturcutre \cite{zno,tsukazaki,falson} the LL gap is
very small. The ratio $\kappa$ between the Coulomb interaction and the LL gap is the relevant
parameter in this context. In GaAs, $\kappa=2.5/\sqrt{B}$, which would be very small 
in a strong magnetic field. In the ZnO heterointerface, where the dielectric constant is $8.5$, that 
ratio is $\kappa=25.1/\sqrt{B}$, i.e., about ten times larger than that of GaAs. Therefore, considering the 
electron system in a single LL may not be appropriate. On the other hand, in graphene the ratio
depends only on the dielectric constant of the substrate \cite{peterson}. In the case of boron nitride 
as the substrate, $\kappa =0.5\sim 0.8$, which is smaller than one. Hence, a perturbative scheme 
of the effective Coulomb potential \cite{bishara}, in which 
higher LLs are projected onto the lowest Landau level by expanding the Coulomb potential in
order of $\kappa$ can be useful. These theories are only useful when $\kappa$ is comparable to or 
smaller than unity. In ZnO, this ratio is usually much larger than 1, even an order of magnitude higher 
than 1. 

Here we introduce another method to project the higher empty LLs onto the relevant LL by 
the virtual process between the empty LLs and full (or partly occupied) LLs. The Coulomb potential is 
screened by all the electrons below the Fermi level. The dielectric constant is then replaced by the 
dielectric function of the momentum. The screened Coulomb potential is calculated in the random phase
approximation (RPA) \cite{aleiner}, and is useful for any $\kappa$. When the LL gap is 
infinitely large the screened Coulomb potential returns to the original (unscreened) value. 
This form of screened Coulomb interaction was used earlier in higher LLs \cite{fogler} and in the
case of skyrmions \cite{luo} in the Hartree-Fock approximation. We use this screened Coulomb interaction 
to study the collective modes of the FQHE states in the ZnO system using the exact 
diagonalization scheme. 

In our scheme for the screened Coulomb potential \cite{luo}, the interaction between electrons in the 
relevant LL is renormalized by the polarizability of all the 
other Landau levels. We consider here only the static screening so that only the zero-frequency 
response function is taken into consideration. The Coulomb potential in the momentum space is
$V\left(\mathbf{q}\right) =\frac{2\pi e^2}{\epsilon q}$. The screened Coulomb potential is then written
$$V^{}_s\left( \mathbf{q}\right) =\frac{2\pi e^2}{\epsilon \epsilon^{}_s\left(\mathbf{q}\right)q},$$
where $\epsilon^{}_s\left(\mathbf{q}\right)$ is the screened dielectric function \cite{shizuya},
$$\epsilon^{}_s\left( \mathbf{q}\right) =1-V\left( \mathbf{q}\right) \chi_{nn}^R\left(\mathbf{q,}
\,\omega\rightarrow0^+\right),$$
$\chi_{nn}^R$ is the retarded density-density response function and the associated response function 
$\chi^{}_{nn}$ is defined as
\[
\chi^{}_{nn}\left(\mathbf{q,}\tau \right)=-\frac1{\hbar S}\left\langle T^{}_{\tau}\delta n\left( 
\mathbf{q,}\tau \right) \delta n\left( -\mathbf{q,} 0\right) \right\rangle, \]
with time ordering operator $T^{}_{\tau }$, system area $S$ and the density operator $n\left( \mathbf{q}
\right)$. If we consider only the non-interacting response function $\chi_{nn}^{0}$ without LL mixing 
in the Matsubara frequency $\Omega^{}_n$, then
$$\chi_{nn}^0\left( \mathbf{q,}i\Omega^{}_n\right)=\frac{N^{}_s}{\hbar S}
\sum^{}_{\sigma ,n,n^{\prime }}\left\vert F^{}_{n^{\prime},n}\left( \mathbf{q}
\right) \right\vert ^{2}\frac{\nu ^{}_{\sigma ,n}-\nu ^{}_{\sigma ,n^{\prime }}}{
i\Omega ^{}_{n}+\left( E^{}_n-E^{}_{n^{\prime }}\right)/\hbar},$$
where $N^{}_s$ is the LL degeneracy, $\sigma$ is the spin index, $n,n^{\prime }$ are the LL indices, 
$E^{}_n$ is the kinetic energy of the LL $n$, and the form factor is defined by
\begin{eqnarray*}
F^{}_{n,n^{\prime}}\left(\mathbf{q}\right) &=&\frac{\sqrt{\min \left(
n,n^{\prime}\right)!}}{\sqrt{\max\left(n,n^{\prime}\right)!}}
e^{-q^2\ell^2/4}L_{\min\left( n,n^{\prime }\right)}^{\left\vert
n-n^{\prime}\right\vert}\left(\frac{q^2\ell^2}2\right) \nonumber \\
&&\times \left[\frac{\left(\mathtt{sign}\left( n-n^{\prime}\right)
q^{}_y+iq^{}_x\right)\ell}{\sqrt2}\right]^{\left\vert n-n^{\prime}\right\vert}
\end{eqnarray*}
with a Laguerre function $L(x)$. The parameter $\nu^{}_{\sigma,n}$ is the filling factor of the level 
with spin $\sigma$ in the LL $n$. In our exact diagonalization scheme $\nu =N^{}_e/N^{}_{\phi}$, 
where $N^{}_e$ is the electron number of the finite-size system.

In order to study the collective modes for odd- and even-denominator FQHE states, we follow the
standard procedure of finite-size systems in a periodic rectangular geometry \cite{book,haldane}.
The Hamiltonian for the Coulomb interaction is
\begin{eqnarray*}
H^{}_C &=&\frac12\sum^{}_{\alpha,\beta}\sum^{}_{n^{}_1,n^{}_2,n^{}_3,n^{}_4}
\sum^{}_{i^{}_1,i^{}_2,i^{}_3,i^{}_4}V_{i^{}_1,i^{}_2,i^{}_3,i^{}_4}^{n^{}_1,n^{}_2,n^{}_3,n^{}_4}
\nonumber \\
&&\times c_{\alpha,n^{}_1,i^{}_1}^{\dag}c_{\beta,n^{}_2,i^{}_2}^{\dag
}c^{}_{\beta,n^{}_3,i^{}_3}c^{}_{\alpha,n^{}_4,i^{}_4},
\end{eqnarray*}
where $n^{}_i$ is the LL index, $i^{}_j$ is the guiding center index, $\alpha,\beta$ 
are spin indices, and $c$ is the electron operator. The Coulomb
interaction elements are given by \cite{yoshioka}
\begin{eqnarray*}
V_{i^{}_1,i^{}_2,i^{}_3,i^{}_4}^{n^{}_1,n^{}_2,n^{}_3,n^{}_4} &=&\frac1{N^{}_s}
\frac{e^2}{\epsilon\ell}\overline{\sum^{}_{\mathbf{q}}}\frac1{\left(
\epsilon^{}_s\right)q\ell}\delta_{i^{}_1,i^{}_4+q^{}_y\ell^2}^{\prime
}\delta _{i^{}_2,i^{}_3-q^{}_y\ell^2}^{\prime} \nonumber \\
&&\times e^{iq^{}_x\left(i^{}_3-i^{}_1\right)}F^{}_{n^{}_1,n^{}_4}\left(\mathbf{q}
\right) F^{}_{n^{}_2,n^{}_3}\left(-\mathbf{q}\right),
\end{eqnarray*}
where $\overline{\sum}$ excludes the term of $\mathbf{q=0}$, $\delta^{\prime }$ includes 
the periodic boundary condition, and the momentum is discrete $\mathbf{q}=\left( 
\frac{2\pi}{L^{}_x}i,\frac{2\pi}{L^{}_y} j\right) $ with the sample length $L^{}_x$ 
and width $L^{}_y$. If a screened Coulomb interaction is taken into consideration, 
we just need to add the dielectric function $\epsilon^{}_s$ in the denominator. The 
classical interaction term in the Hamiltonian which is induced by the periodic
geometry is neglected even in the screened case, since the term is always a constant.

In the present case of ZnO the Zeeman energy ($0.2489B$ meV) is very close to the LL gap
($0.26311B$ meV). For example, the level $\left\vert 1,\uparrow \right\rangle $ is only a 
little higher than $\left\vert 0, \downarrow \right\rangle $. For odd denominator FQHE, 
for simplicity and without loss of generality, we consider only one LL and compare the 
collective modes with and without screening for filling factors $\nu =k/3$, since the 
spin is polarized. This work focuses on the even denominator FQHE \cite{falson}. In a 
perpendicular magnetic field, $\nu =3/2$ state is not observed as is the case in GaAs 
system. Electrons in the half filled level $\left\vert 0,\downarrow \right\rangle$ 
is compressible. In a tilted field there is a crossover of kinetic energies between 
LL 1 and LL 0 with different spins. The exact diagonalization in a tilted magnetic field
is quite involved \cite{tilted} and is beyond the scope of this paper.

\begin{figure}
\includegraphics[width=7.0cm]{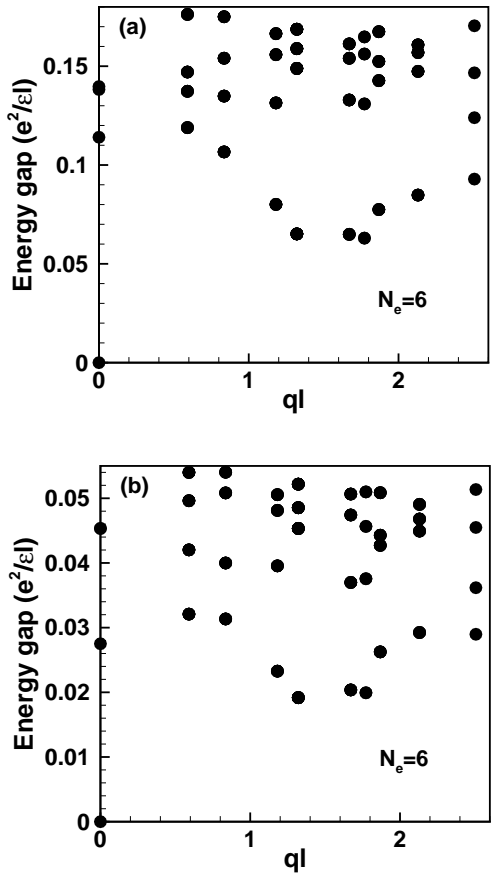}
\caption{\label{figure1} 
The collective mode of $\protect\nu =1/3$ for six electrons, (a) without and (b) with screening.
}\end{figure}    

As mentioned above, in the experiment of \cite{falson} there is no indication of the $\frac52$ state, which is
quite strong in the GaAs system. There could be several possible reasons for this: (i) the LL mixing 
may decrease or even close the gap of the incompressible ground state; (ii) a spin-mixed charge density 
wave state may exist between $\left\vert 0,\downarrow \right\rangle $ and $\left\vert 1,\uparrow 
\right\rangle,$ since the gap $\Delta $\ between the two levels is very small (for $B=3.75$T, the gap 
is only $\Delta =0.05329$ meV $=0.004167e^{2}/\epsilon\ell$ \cite{falson}); or (iii) the screened 
Coulomb potential which integrates out all other LLs, changes the ground state. To test the first 
possibility we perform an exact diagonalization with LL mixing which includes LL $\left\vert 1,\uparrow 
\right\rangle $ and $\left\vert 2,\uparrow \right\rangle $. The results indicate that the collective modes 
are just slightly changed and the ground state is still an incompressible liquid. The spin remains
fully polarized in our numerical calculations that includes $\left\vert 1,\uparrow \right\rangle $ and 
$\left\vert 1,\downarrow \right\rangle $, as in previous theoretical works \cite{TC_even} and
in some of the experimental works \cite{muraki}. On the other hand, if the LL mixing or spin mixing change the ground 
state at $5/2$, then the incompressible ground state at $7/2$ would also be changed. But the FQHE
experiment shows a robust $\nu=7/2$. To test the second possibility, we also perfom an exact diagonalization 
calculation where we class the Hamiltonian by the spin polarization \cite{tapash2,yoshioka3}.
The ground state always has all electrons occupied in $\left\vert 0,\downarrow \right\rangle $ when the 
gap $\Delta\geq 0$. Even for a negative gap $\Delta^{}_C<\Delta <0$, i.e., $\left\vert 1,\uparrow
\right\rangle $ is a little lower than $\left\vert 0,\downarrow \right\rangle $, the electrons of the 
ground state are still in $\left\vert 0,\downarrow \right\rangle $. Note that $\Delta^{}_C$ can not be too
negative: if $\Delta^{}_C\rightarrow -\infty $, then all electrons would be
flipped to $\left\vert 1,\uparrow \right\rangle $.

\begin{figure}
\includegraphics[width=7.0cm]{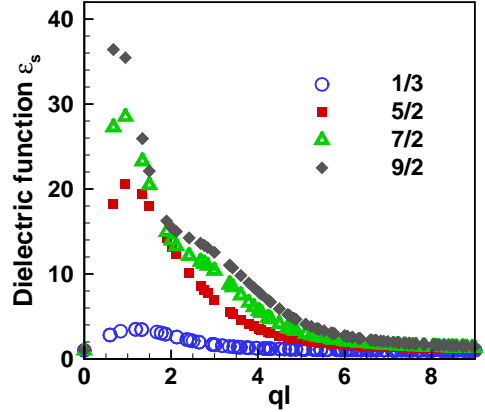}
\caption{\label{figure2}
The dielectric function $\protect\epsilon^{}_{s}$ for filling factors $\protect\nu =1/3,5/2,7/2,9/2$.
}\end{figure}

Only the third possibility seems to explain the experiment, i.e., the absence of the $5/2$ state, while 
appearence of $\nu=7/2$. For simplicity, we consider only a single LL with the screened potential. 
In our work that follows, the aspect ratio is $L^{}_x/L^{}_y=1$. For simpilicity, we only consider the 
filling factors $\nu =k/3,$ $k=1,2,4,5,7,8,10,11.$ Figure \ref{figure1} shows the comparison of the 
unscreened and screened collective modes at $\nu =1/3$ for six electrons. The shape of the characteristic 
FQHE collective mode does not change, only the gap is reduced by the screening. For other filling factors, 
$\nu =k/3,$ $\left( k=2,4,5,7,8,10,11\right),$ 
we are also able to observe the characteristic FQHE collective modes, and the ground states indicate the 
incompressible liquid phase. Without screening, the collective modes in the exact diagonalization are 
calculated in GaAs in Ref.~\cite{TC_even}, where the screening effect is much weaker than for the ZnO 
heterojunction. First, we use the system parameters of GaAs to perform the exact diagonalization with
screened Coulomb potential, and it shows that the FQHE is able to survive for both $5/2$ and $7/2$. It proves 
that our screening calculations are compatible with the GaAs systems. For the ZnO system, we adopt the 
experimental parameters of Ref.~\cite{falson}. The dielectric functions for $\nu =5/2$ and $7/2$ are 
indicated in Fig.~\ref{figure2}. The $7/2$ and $5/2$ are equivalent without screening due to the
electron-hole symmetry in LL $n=1$. The Coulomb interactions are distinguishable with screening included: the 
screening at $7/2$ is stronger than that at $5/2$, and there is an obvious step in the curve at $7/2$. So 
the ground state and collective modes can be different in the two cases.

\begin{figure}
\includegraphics[width=7.0cm]{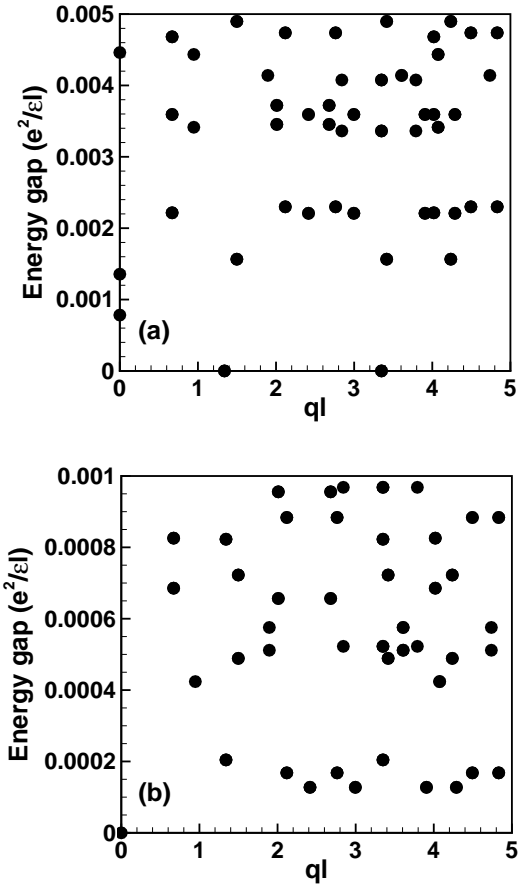}
\caption{\label{figure3}
The exact diagonalization of a $N^{}_e=7$ electrons system. (a) The collective mode for 5/2: the 
ground state is degenerate and compressible. (b) The collective mode for 7/2 indicates an
imcompressible ground state.
}\end{figure}

We have tested different system sizes: $N^{}_e=4\ldots 11$. For simplicity, only the case of $N^{}_e=7$ is 
shown in Fig.~\ref{figure3}. Clearly, the FQHE state is absent for $5/2$, but survives at $7/2$ , even
though the screening of the latter is stronger. The ground state of $5/2$ is a degenerate compressible state, 
but the ground state of $7/2$ is always an incompressible state. Note that for odd electrons, the ground 
states of $7/2$ are at $\mathbf{q=0}$, but for even electrons, the ground states are always located at 
$\mathbf{q}=\sqrt{2\pi /N^{}_s}\left( N/2,N/2\right)$. So the ground state could become an incompressible 
liquid state by a global translation, which was already pointed out in Ref.~\cite{TC_even}.
The collective modes at $7/2$ seem to have two minimum that are located at about $q\ell =2.5$ and $3.8$. The 
energy gap, however, is very small compared to other systems. It is because the screened Coulomb 
intearction suppresses the gap. Interestingly, the screening of $7/2$ is stronger, but the FQHE is still
not destroyed. The energy gap for a larger system (more electrons) is larger than that of a smaller system 
(for example, when $N^{}_e=11$, the lowest gap is 0.0004$e^2/\epsilon\ell$). So we expect that for a real 
system, the energy gap is large enough to be observable. 

\begin{figure}
\includegraphics[width=7.0cm]{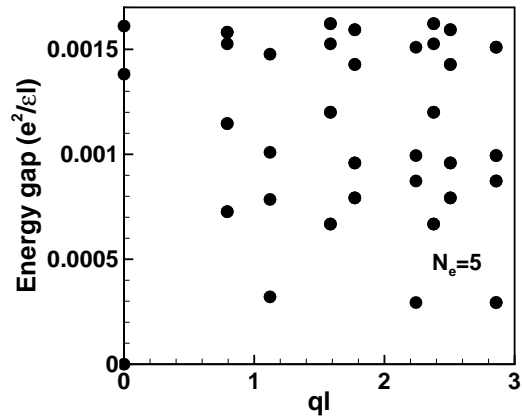}
\caption{\label{figure4}
The exact diagonalization results with screening for a $N_e=5$ system at $\nu=9/2$. 
}\end{figure}

For higher LLs, at $\nu = 9/2$, $\kappa$ is even larger than that in LL 1 and the screening is stronger. 
The Coulomb potential thus be changed more by the screening induced by other LLs. Our exact diagonalization 
results are presented for 5 electrons at the experimental value of $B=2.1$ T. The collective modes clearly
show an incompressible state (Fig.~\ref{figure4}). However, the gap is very small. Incidentally, the experimental 
signal is also very weak. 

To summarize, we have studied the FQHE states in the ZnO system with screened Coulomb interaction
that incorporates the influence of other landau levels. For the odd-denominator filling
factors, our work agrees with the present system of ZnO and with earlier GaAs systems as well. However,
for the even-denominator filling factors, we are able to explain the absence of $3/2,5/2$ FQHE states,
but the presence of $7/2,9/2$ FQHE states, by introducing screening which integrates out all the other 
LLs. The screening discussed in this paper is only the static one, which means that we run the risk
overscreening the Coulomb interaction. The dynamic screening may screen the Coulomb potential weakly. 
However, we expect that the results obtained here would not essentially change.

The work has been supported by the Canada Research Chairs Program of the Government of Canada.
W. Luo would like to thank R. C\^{o}t\'{e} for helpful discussions and Huizhong Lu of Calcul Qu\'ebec 
for help with computations. The computation time was provied by Calcul Qu\'ebec and Compute Canada.

\end{document}